\documentclass[manuscript,authorversion,review=false,timestamp=false]{acmart}

\usepackage{balance}

\begin{document}



\title[Implications of Regulations on the Use of AI and Generative AI for HCR-AI]{Implications of Regulations on the Use of AI and Generative AI for Human-Centered Responsible Artificial Intelligence}

\author{Marios Constantinides}
\orcid{0000-0003-1454-0641}
\affiliation{
  \institution{Nokia Bell Labs}
  \city{Cambridge}
  \country{UK}
}
\email{marios.constantinides@nokia-bell-labs.com}

\author{Mohammad Tahaei}
\authornote{Also affiliated with eBay.}
\orcid{0000-0001-9666-2663}
\affiliation{
  \institution{International Computer Science Institute}
  \state{CA}
  \country{USA}
}
\email{mtahaei@icsi.berkeley.edu}

\author{Daniele Quercia}
\orcid{0000-0001-9461-5804}
\affiliation{
  \institution{Nokia Bell Labs}
  \city{Cambridge}
  \country{UK}
}
\email{daniele.quercia@nokia-bell-labs.com}

\author{Simone Stumpf}
\orcid{0000-0001-6482-1973}
\affiliation{
    \institution{University of Glasgow}
    \city{Glasgow}
    \country{UK}
}
\email{simone.stumpf@glasgow.ac.uk}

\author{Michael Madaio}
\orcid{0000-0003-2133-0810}
\affiliation{
    \institution{Google}
    \city{}
    \country{USA}
}
\email{madaiom@google.com}

\author{Sean Kennedy}
\orcid{0000-0003-0000-1786}
\affiliation{
  \institution{Nokia Bell Labs}
  \city{Ottawa}
  \country{Canada}
}
\email{sean.kennedy@nokia-bell-labs.com}

\author{Lauren Wilcox}
\orcid{0000-0001-6598-1733}
\affiliation{
    \institution{eBay}
    \city{San Jose}
     \state{CA}
    \country{USA}
}
\email{lgw231@acm.org}

\author{Jessica Vitak}
\orcid{0000-0001-9362-9032}
\affiliation{
    \institution{University of Maryland, College Park}
    \state{MD}
    \country{USA}
}
\email{jvitak@umd.edu}

\author{Henriette Cramer}
\orcid{0000-0002-0786-0324}
\affiliation{
    \institution{Independent}
    \city{San Francisco}
    \country{USA}
}
\email{henriette.cramer@gmail.com}

\author{Edyta Bogucka}
\orcid{0000-0002-8774-2386}
\affiliation{
  \institution{Nokia Bell Labs}
  \city{Cambridge}
  \country{UK}
}
\email{edyta.bogucka@nokia-bell-labs.com}

\author{Ricardo Baeza-Yates}
\orcid{0000-0003-3208-9778}
\affiliation{
    \institution{EAI, Northeastern University}
    \state{CA}
    \country{USA}
}
\email{rbaeza@acm.org}

\author{Ewa Luger}
\orcid{0000-0001-7882-9415}
\affiliation{
    \institution{University of Edinburgh}
    \city{Edinburgh}
    \country{UK}
}
\email{ewa.luger@ed.ac.uk}

\author{Jess Holbrook}
\orcid{0000-0001-9398-6056}
\affiliation{
    \institution{Meta}
    \city{Seattle}
    \country{USA}
}
\email{jess.holbrook@gmail.com}

\author{Michael Muller}
\orcid{0000-0001-7860-163X}
\affiliation{
    \institution{IBM Research AI}
    \city{Cambridge}
    \state{MA}
    \country{USA}
}
\email{michael_muller@us.ibm.com}

\author{Ilana Golbin Blumenfeld}
\orcid{0000-0001-8366-8468}
\affiliation{
    \institution{PwC}
    \city{Los Angeles}
    \state{CA}
    \country{USA}
}
\email{ilana.a.golbin@pwc.com}

\author{Giada Pistilli}
\orcid{0000-0003-4941-0505}
\affiliation{
    \institution{Hugging Face}
    \city{Paris}
    \state{}
    \country{France}
}
\email{giada@huggingface.co}

\renewcommand{\shortauthors}{Constantinides et al.}

\begin{abstract}
With the upcoming AI regulations (e.g., EU AI Act) and rapid advancements in generative AI, new challenges emerge in the area of Human-Centered Responsible Artificial Intelligence (HCR-AI). As AI becomes more ubiquitous, questions around decision-making authority, human oversight, accountability, sustainability, and the ethical and legal responsibilities of AI and their creators become paramount. Addressing these questions requires a collaborative approach. By involving stakeholders from various disciplines in the 2\textsuperscript{nd} edition of the HCR-AI Special Interest Group (SIG) at CHI 2024, we aim to discuss the implications of regulations in HCI research, develop new theories, evaluation frameworks, and methods to navigate the complex nature of AI ethics, steering AI development in a direction that is beneficial and sustainable for all of humanity.

\end{abstract}




\maketitle

\section{Motivation \& Background}
In 2023, the first Human-Centered Responsible Artificial Intelligence (HCR-AI) Special Interest Group (SIG) was held at the ACM CHI 2023 conference. This SIG is a continuation of the themes explored in previous CHI workshops between 2020 and 2022~\cite{lee2020human, ehsan2021operationalizing, ehsan2022human}, and it aims to bring together researchers and practitioners from both academia and industry to map current and future research trends in HCR-AI~\cite{tahaei2023human}. Broadly defined, HCR-AI champions the development of Artificial Intelligence (AI) systems that not only benefit individuals and societies but also proactively identify and mitigate potential harms through the lens of human-centered design. Although the significance of the human element in AI has been acknowledged for some time---as evidenced by the early works of Friedman~\cite{friedman1966bias} and Suchman~\cite{suchman1987plans}---it is only in recent years that the field of human-centered computing research has significantly engaged with integrating people and their values into the design and development of AI systems~\cite{aragon2022human, shneiderman2022human, tahaei2023toward}. 

The 1\textsuperscript{st} edition of the HCR-AI SIG was marked by collaborative efforts and led to a comprehensive list of ideas and topics. Discussions covered a wide array of subjects, from tools and frameworks for responsible AI design and development to fairness and explainability, as well as the integration of Human-Computer Interaction (HCI) theories into AI research. In the 2\textsuperscript{nd} edition, given the rapid advancements in generative AI and large language models, we aim to cover pressing issues such as use of AI regulations, governance frameworks, accountability, sustainability, and the ethical and legal responsibilities of AI and its creators. Our goal is to establish a platform for discussing the implications of regulations on HCI and AI research, develop new theories, evaluation frameworks, and methodologies to effectively navigate the fast-moving landscape of AI research, informed by regulatory efforts. Below, we highlight a selection of studies that exemplify the issues revolving around the implications of use of AI regulations and frameworks, all of which are relevant to the focus of the Special Interest Group (SIG). It is important to note that this list is not exhaustive, but rather serves to illustrate the breadth and depth of the existing body of work.

The landscape of use of AI regulation and governance is continually evolving, marked by significant developments. Key among these developments are the proposal of the AI Act in the European Union~\cite{eu_ai_act_2022}, the Executive Order issued by US President Biden on October 30, 2023, focusing on the Safe, Secure, and Trustworthy Development and Use of AI~\cite{us_ai_bill_order}, the G7's establishment of a Code of Conduct for entities creating advanced AI systems~\cite{g7}, and the development of the AI Risk Management Framework by the US National Institute of Standards and Technology (NIST)~\cite{nist2023aiRisk}. At a conceptual level, all governance efforts share some common themes such as ensuring that AI systems introduced to the market are safe, respect human rights, and comply with the law. While specific approaches to AI risk management and mitigation may vary, the debate has intensified, especially following the introduction of large language models (e.g., ChatGPT), regarding how to regulate and use these models responsibly. In particular, our community needs to stay informed and understand the implications of these regulations, as well as how they apply in research and practice. For example, the ACM Technology Policy Council (TPC) has released a statement on principles for responsible algorithmic systems, laying out nine instrumental principles intended to foster fair, accurate, and beneficial algorithmic decision-making~\cite{acm_principles_2022}; these principles have also been extended for the development, deployment, and use of generative AI technologies~\cite{acm_principles_2023}.

Developing governance regimes for AI systems is crucial for making AI accountable, fair, and transparent~\cite{cath2018governing}, as well as for preventing the unsustainable large carbon footprint associated with training these systems~\cite{thompson2021deep, strubell2019energy, bender2021dangers}. AI's impact on privacy and the potential replication of bias and discrimination highlight the need for ethical AI development and deployment~\cite{raso2018artificial}. To facilitate research and development of responsible, explainable, and ethical AI for an inclusive society, it is important to adopt a process-oriented approach for regulating the use of AI systems~\cite{stranieri2022process, iaaa2023}. 

A growing body of research---typically discussed in conferences with a long-standing commitment to human-centered design, such as the Conference on Human Factors in Computing Systems (CHI), the Conference on Computer-Supported Cooperative Work and Social Computing (CSCW) and Intelligent User Interfaces (IUI), as well as in newer conferences like the Conference on AI, Ethics, and Society (AIES) and the Conference on Fairness, Accountability, and Transparency (FAccT)---focuses on providing practical tools and frameworks to enable AI researchers and practitioners addressing responsible AI issues. For example, tools and frameworks have been proposed to assist developers in mitigating biases~\cite{bird2020fairlearn, gebru2021datasheets}, explaining algorithmic decisions~\cite{arya2019one}, and ensuring privacy-preserving AI systems~\cite{fjeld2020principled}. However, these toolkits are often designed to offer technical solutions to broader sociotechnical challenges. A growing body of literature is attempting to understand how existing responsible AI toolkits align with HCI practitioners' practices, revealing a number of shortcomings such as a disconnect between the needs of practitioners and the tools offered by fairness research~\cite{richardson2021towards}. 
Kaur et al. found that data scientists often struggle to understand the visualizations from interpretability tools such as InterpretML~\cite{interpret_ml} and SHAP~\cite{lundberg_unified_2017}. Balayn et al.~\cite{balayn2023fairness} observed that less experienced practitioners tend to use limited metrics and methods in fairness toolkits, while those with more experience and interdisciplinary backgrounds seek more innovative solutions. Similarly, 
without standardized methods to assess fairness in recommendation systems, industry practitioners, often non-experts, face challenges or adopt inadequate metrics for their applications~\cite{beattie2022challenges, smith2023scoping}. Nakao et al. \cite{nakao_toward_2022} have proposed interfaces to involve lay users in fairness assessment and mitigation. Deng et al.~\cite{deng2022exploring} noted the lack of standardized guidelines in fairness toolkits for explaining fairness issues to non-technical team members. Yildirim et al.~\cite{yildirim2023investigating} reported a desire among AI practitioners for toolkits that aid in communication, suggesting the inclusion of design notes with suggestions and best practices, while Wang et al. ~\cite{wang2023designing} found that industry practitioners need support in three key areas: building and reinforcing a ``responsible lens'' on AI, prototyping AI applications responsibly, and conducting multi-stage evaluations of AI applications.
These studies collectively highlight the need for more human-centered AI toolkits and processes to support practitioners in various aspects of AI development and communication; a discussion we foresee to cover during the SIG.

Another crucial topic in the public debate is that of sustainability. With the increasing power consumption needed to train AI systems, sustainability is becoming a critical focus area. This involves considering the environmental impact of AI systems, such as the energy consumption of data centers and the carbon footprint of training large models~\cite{strubell2019energy}. The need for sustainable AI practices extends beyond environmental concerns to encompass long-term viability and responsible resource usage in AI development and deployment. Research and discussions in this area emphasize the importance of creating AI systems that are not only efficient and effective but also minimize ecological harm~\cite{rolnick2022tackling}. Additionally, sustainable AI involves ensuring that AI systems contribute positively to societal goals, aligning with sustainable development objectives (e.g., United Nations Sustainable Development Goals)~\cite{vinuesa2020role}. As AI continues to evolve, incorporating sustainability into AI regulations and governance frameworks becomes essential, guiding researchers and practitioners towards environmentally conscious and socially responsible AI solutions.

\section{HCR-AI SIG at CHI 2023}
The SIG at CHI 2023~\cite{tahaei2023human} saw a turnout of over 100 diverse participants, forming 12 groups that discussed various topics such as tools, use cases, and associated risks. All collaborative efforts were documented using Google Slides and Miro boards, with each group presenting their ideas followed up by a Q\&A from the audience. The outcomes were shared with the community via a dedicated website\footnote{HCR-AI website: \url{https://hcrai.github.io/}; the site will also host future HCR-AI editions.}, extending the reach to participants and non-participants. In the inaugural edition, our objective was to navigate the expansive landscape. Building on that experience, this year's focus shifted to specific areas of interest, garnering increased attention at CHI 2023 and aligning with the latest developments in AI. Staying true to our commitment outlined in the CHI 2023 proposal, we are dedicated to evolving this SIG into a recurring series. With a diverse mix of organizers from last year's SIG and new organizers, our aim remains steadfast---to assemble a diverse audience for meaningful discussions on HCR-AI.

\section{Proposal \& SIG's Goal at CHI 2024}
This SIG is a continuation of the CHI 2023 SIG~\cite{tahaei2023human}. The evolving landscape of AI makes the timing ideal for a SIG at CHI 2024. We anticipate that a 2\textsuperscript{nd} edition of the HCR-AI will benefit the CHI community, fostering a wider network of researchers engaging with the pressing issues of human-centered responsible AI research. We anticipate participants across industry and academia, from various career stages, bringing diverse expertise from fields such as theoretical computer science, social computing, machine learning, human-computer interaction, and social sciences and humanities. Gathering these experts in a combined physical and virtual setting for 75 minutes will be valuable for community building and brainstorming, facilitating the creation of a trend map in this area through activity diagramming. We plan to use digital platforms such as Miro, Slack, and Google Slides to document the collective knowledge generated, maintain ongoing communication within the CHI community, and include remote participants effectively.

\section{Expected Outcomes \& Next Steps}
We plan to distribute the Miro board to participants and make it accessible to the public, aiding further research in HCR-AI. Additionally, a dedicated Slack channel will be established for ongoing dialogue. The primary aim of the SIG is to cultivate a community among HCR-AI researchers from both academic and industrial backgrounds, encouraging collaborative efforts. This SIG presents a unique chance to unite individuals interested in HCR-AI at CHI, fostering a strong network within this field.

Following the SIG, we intend to hold virtual meetings every six months with the attendees to discuss new ideas and recent developments. We will also set up a website to disseminate the outcomes from the SIG. Given the popularity of the topic, it is important to recognize that groups beyond the CHI community, which are dedicated to implementing responsible AI in practice, could benefit from a wider dissemination of the SIG's outcomes. We intend to engage with external communities such as those involved in responsible product and innovation~\cite{ten2022responsible} as well as recommendation decisions, which are closely linked to trust and safety strategies, through blog posts and LinkedIn and X (formerly Twitter) posts to extend our reach. Additionally, the co-organizing team consists of researchers from both academia and industry who will help spread the word through their professional networks, and ideally attract participants with an AI governance perspective. Efforts will also be made to encourage participants to pursue joint funding opportunities, collaborate on research papers, and to consider the feasibility of organizing symposiums and events for knowledge sharing.

\balance
\bibliographystyle{ACM-Reference-Format}
\bibliography{biblio}

\end{document}